\documentclass[aps,prapplied,twocolumn,longbibliography]{revtex4-1}
\usepackage{epsf,graphicx}
\usepackage{amssymb}
\usepackage{amsmath}
\usepackage{bm}
\usepackage{epstopdf}
\usepackage{xcolor}
\usepackage{hyperref}
\usepackage{float}
\usepackage{lineno,hyperref,amsmath}
\usepackage{booktabs}
\usepackage{graphicx}
\usepackage{subfigure}
\usepackage{float}
\usepackage{amsmath}
\usepackage{amssymb}
\usepackage{appendix}
\usepackage{tikz}
\usepackage{soul}

\hypersetup{hidelinks,
	colorlinks=true,
	allcolors=black,
	pdfstartview=Fit,
	breaklinks=true}

\definecolor{lime}{HTML}{A6CE39}
\DeclareRobustCommand{\orcidicon}{
	\begin{tikzpicture}
		\draw[lime, fill=lime] (0,0)
		circle[radius=0.16]
		node[white]{{\fontfamily{qag}\selectfont \tiny \.{I}D}};
	\end{tikzpicture}
	\hspace{-2mm}
}
\foreach \x in {A, ..., Z}{%
	\expandafter\xdef\csname orcid\x\endcsname{\noexpand\href{https://orcid.org/\csname orcidauthor\x\endcsname}{\noexpand\orcidicon}}
}

\begin{document}
	\title{Quantum Transport Reservoir Computing}
	\author{Yecheng Jing \hspace{-1.5mm}\orcidB{}}
	\thanks{The authors contribute equally.}
	\affiliation{National Laboratory of Solid State Microstructures, School of Physics, 
		Jiangsu Physical Science Research Center, and Collaborative Innovation Center of Advanced Microstructures,
		Nanjing University, Nanjing 210093, China}
	\author{Pengfei Wang}
	\thanks{The authors contribute equally.}
	\affiliation{National Laboratory of Solid State Microstructures, School of Physics, 
		Jiangsu Physical Science Research Center, and Collaborative Innovation Center of Advanced Microstructures,
		Nanjing University, Nanjing 210093, China}
	\author{Shuai Zhang}
	\affiliation{National Laboratory of Solid State Microstructures, School of Physics, 
		Jiangsu Physical Science Research Center, and Collaborative Innovation Center of Advanced Microstructures,
		Nanjing University, Nanjing 210093, China}
	\author{Zhoujie Zeng}
	\affiliation{National Laboratory of Solid State Microstructures, School of Physics, 
			Jiangsu Physical Science Research Center, and Collaborative Innovation Center of Advanced Microstructures,
			Nanjing University, Nanjing 210093, China}
	\author{Shi-Jun Liang}
	\email{Corresponding author: sjliang@nju.edu.cn}
	\affiliation{National Laboratory of Solid State Microstructures, School of Physics, 
		Jiangsu Physical Science Research Center, and Collaborative Innovation Center of Advanced Microstructures,
		Nanjing University, Nanjing 210093, China}
	\author{Wei Chen \hspace{-1.5mm}\orcidA{}}
	\email{Corresponding author: pchenweis@gmail.com}
	\affiliation{National Laboratory of Solid State Microstructures, School of Physics, 
		Jiangsu Physical Science Research Center, and Collaborative Innovation Center of Advanced Microstructures,
		Nanjing University, Nanjing 210093, China}
	\date{\today}
	
	\begin{abstract}
		Reservoir computing (RC), a neural network designed for temporal data, enables efficient computation with low-cost training and direct physical implementation. Recently, quantum RC has opened new possibilities for conventional RC and introduced novel ideas to tackle open problems in quantum physics and advance quantum technologies. Despite its promise, it faces challenges, including physical realization, output readout, and measurement-induced back-action. Here, we propose to implement quantum RC through quantum transport in mesoscopic electronic systems. Our approach possesses several advantages: compatibility with existing device fabrication techniques, ease of output measurement, and robustness against measurement back-action. Leveraging universal conductance fluctuations, we numerically demonstrate two benchmark tasks, spoken-digit recognition and time-series forecasting, to validate our proposal. This work establishes a novel pathway for implementing on-chip quantum RC via quantum transport and expands the mesoscopic physics applications.
	\end{abstract}
	
	\maketitle
	\section{Introduction}
\begin{figure}[tb]
		\begin{center}
			\includegraphics[width=\columnwidth]{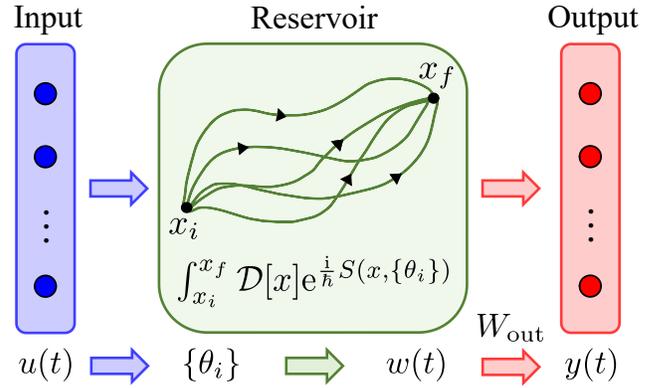}
		\end{center}
		\caption{Architecture of a new quantum RC type. The input $u(t)$ is initially encoded in the reservoir. By adjusting parameters $\{\theta_i\}$ of the quantum system action $S$ using $u(t)$, the quantum system produces the internal state $w(t)$ nonlinearly. This process maps $u(t)$ nonlinearly into a high-dimensional space. The weight matrix $W_{\text{out}}$ is subsequently optimized linearly to produce the final $y(t)$.}
		\label{RC}
	\end{figure}

	Reservoir computing (RC) is a neural network-based framework designed to efficiently process time-dependent data by following two steps to achieve the recurrent neural network effect~\cite{jaeger2001echo,maass2002real,jaeger2004harnessing,verstraeten2007experimental,tanaka2019recent,cucchi2022hands,yan2024emerging}, as depicted in Fig.~{\ref{RC}}. 
	Firstly, the temporal input $u(t)$ undergoes a nonlinear mapping into a high-dimensional space through a fixed reservoir, resulting in the internal state $w(t)$. Subsequently, the weight matrix $W_{\text{out}}$ is linearly optimized to acquire the final readout $y(t)$. Compared to the conventional recurrent neural network approach, the utilization of a fixed reservoir significantly reduces training costs and enables physical implementations across a wide range of systems, including electronic~\cite{appeltant2011information,du2017reservoir,moon2019temporal,zhong2021dynamic,zhang2022reconfigurable,gao2024toward,kim2024analog}, photonic~\cite{vandoorne2014experimental,vinckier2015high,larger2017high,rafayelyan2020large,genty2021machine,shastri2021photonics,yaremkevich2023chip}, and spintronic systems~\cite{torrejon2017neuromorphic,prychynenko2018magnetic,romera2018vowel,pinna2020reservoir,yamaguchi2020periodic,watt2021implementing,lee2022reservoir}, among others~\cite{toprasertpong2022reservoir,liu2023interface}. 
	
	Besides extensive explorations in classical systems, recent efforts to realize RC on quantum platforms have also attracted considerable interest. This strategy not only brings novel resources for RC~\cite{fujii2017harnessing,nakajima2019boosting,govia2021quantum,bravo2022quantum,kalfus2022hilbert,sakurai2022quantum,dudas2023quantum,garcia2023scalable} but also introduces innovative methodologies with significant potential for addressing unresolved challenges in quantum physics and technologies~\cite{ghosh2019npj,ghosh2019prl,angelatos2021reservoir}. Meanwhile, several significant challenges stand in the way~\cite{mujal2021opportunities}.
	Firstly, translating some theoretical proposals into physical implementations remains a hurdle. Additionally, extracting readouts---typically the expectation values of physical observables---requires preparing multiple copies of the system and performing repeated measurements. Finally, the measurement process itself can induce back-action to the quantum reservoir, disrupting its intended behavior. While several physical implementations of quantum RC exist, addressing these challenges remains difficult. For instance, experimental implementations on IBM quantum platforms extract outputs through projective measurements, making the measurement back-action non-negligible~\cite{chen2020temporal,kubota2023temporal,yasuda2023quantum}. At each time point, the reservoirs must undergo a realization cycle: resetting to $t=0$ before being measured at $t=k\Delta t$. 
	For quantum RC realized via superconducting circuits~\cite{senanian2024microwave}, measurement back-action is utilized to generate non-unitary dynamics as a complementary mechanism interspersed throughout the evolution, though this approach requires $10^3-10^4$ measurement repetitions to obtain expectation values as internal states and achieve high accuracy. 
	Consequently, these experiments require multiple system copies and repeated measurements.
	
	To address these challenges, it is insightful to reconsider the basic principles underlying quantum RC. 
	The primary advantage of a quantum reservoir can be well interpreted by the path integral formalism of the quantum propagator, given by
	\begin{equation}
		\begin{split}
			G(x_f, x_i, t)&=\int_{x_i}^{x_f}\mathcal{D}[x(t)] \text{e}^{\frac{\text{i}}{\hbar}S[x(t),\{\theta_i\}]}, \\
			S[x(t),\{\theta_i\}]&=\int_0^t dt'\mathcal{L}\left(x,\dot{x},\{\theta_i\}\right),
		\end{split}
	\end{equation}
	where $S$ denotes the action associated with the classical paths $x(t)$, and $\{\theta_i\}$ represent tunable physical parameters of the 
	system; see Fig.~\ref{RC}. The parameters $\{\theta_i\}$ influence the propagator in a highly nonlinear way due to the interference among an infinite number of paths, a direct consequence of the superposition principle in quantum mechanics. This inherent quantum nonlinearity makes quantum systems an ideal platform for RC, capable of generating outputs with nonlinear dependence on $\{\theta_i\}$. Specifically, by carefully selecting appropriate physical 
	quantities as input and output, the quantum platform can fulfill the core requirements of RC: separation property and short-term memory. The separation property ensures that distinct inputs yield distinguishable outputs, while the short-term memory implies that the reservoir's output depends not only on the current input but also on recent inputs.
	
	The complete architecture of a quantum RC through quantum propagation is illustrated in Fig.~{\ref{RC}}. Manipulating physical parameters $\{\theta_i\}$ with a subset of the temporal input $u(t)$ enables the quantum system to produce an output. The internal state $w(t)$ is then collectively formed by all outputs after the complete use of $u(t)$. Building on this perspective, here, we propose implementing quantum RC through quantum transport in mesoscopic systems. This approach not only serves as a direct manifestation of the quantum propagator but also addresses the challenges encountered in previous studies. The first advantage lies in the established device fabrication techniques and the diverse manipulation and measurement approaches, which significantly enhance the feasibility of its physical implementation. Secondly, easily measurable outputs, such as current, arise from a sufficient number of electrons, effectively enabling the repeated measurement of quantum states without the need for multiple copies of the devices. Finally, the measurement process has little impact on the reservoir, ensuring the system maintains its intended behavior. These advantages establish mesoscopic electron transport as a promising avenue for the implementation of practical quantum RC. 
	
	In this work, we employ a typical but underutilized mesoscopic transport phenomenon of universal conductance fluctuations (UCF) to demonstrate the feasibility and potential of our approach.
	Extending quantum RC to other quantum transport scenarios is straightforward.
	In this scenario, the physical parameters $\{\theta_i\}$ represent the gate voltages applied to the mesoscopic conductor and the internal state $w(t)$ of the reservoir is associated with the transport current. Based on the numerical results, such a quantum reservoir demonstrates competitive performance and precise electrical tunability in two benchmark tasks: spoken-digit recognition and time-series forecasting.
	This work establishes a novel pathway for realizing on-chip quantum RC through quantum transport, addressing challenges encountered in previous studies via its compatibility with existing industrial equipment, easily measurable output, and robustness to measurement induced back-action. Additionally, this work also enriches the application scope of mesoscopic physics.
	
	\begin{figure}[tb]
		\begin{center}
			\includegraphics[width=\columnwidth]{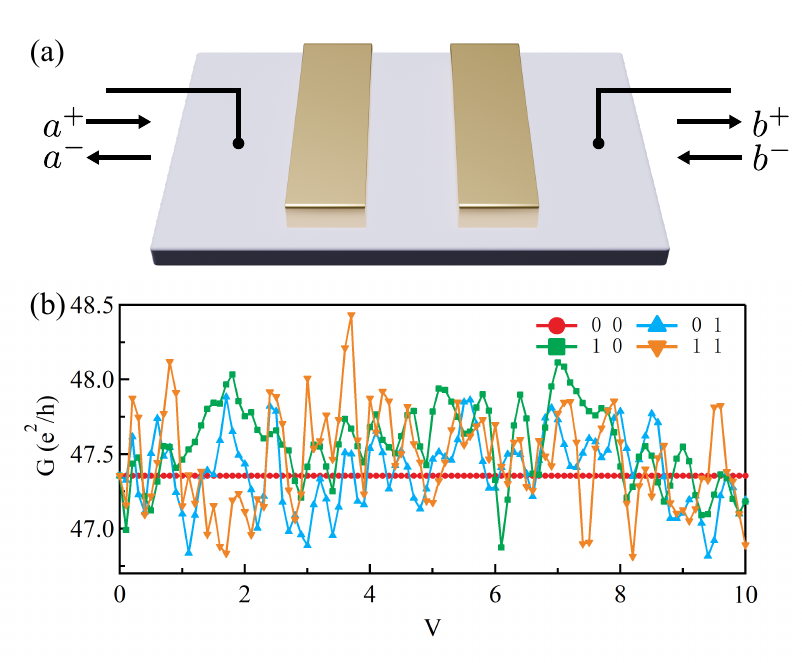}
		\end{center}
		\caption{(a) Schematic of a quantum reservoir using UCF with two gates. Parameters $\{\theta_i\}$ here are various gate voltages applied on the sample, and the internal state $w(t)$ is formed by the resulting current values. (b) UCF induced by varying gate voltages. Encodings `0' and `1' are represented by gate voltages $0$ and $V$, respectively. Various encodings result in distinct conductance values across a wide range. Performance of our reservoir can be precisely tuned via gate voltages. The relevant parameters are: \(M=30\), \(l_{el}=100/3\), \(l=l_{el}\), 10 impurities.}
		\label{Experimental device and G}
	\end{figure}
	
	\section{Model and method}
	\label{section:model and methods}
	We consider a disordered conductor with its size $L$ smaller than the phase coherence length $l_\phi$. In this regime, the conductance $G$ of the sample exhibits 
	reproducible and aperiodic fluctuations in response to changes in gate voltages, magnetic field, or impurity
	configurations, arising from intricate quantum interference effect. Such conductance fluctuations are on the order of $e^2/h$, independent of the sample's geometry or dimensionality, and are referred to as UCF~\cite{umbach1984magnetoresistance,lee1985universal}. This phenomenon has attracted extensive theoretical \cite{altshuler1986repulsion,lee1987universal,blanter1996electron,aleiner2002inelastic,kharitonov2008universal}, numerical \cite{stone1985magnetoresistance,cahay1988conductance,
		rycerz2007anomalously,choe2015universal,hu2017numerical,shafiei2024tailoring}, and experimental studies \cite{webb1985observation,benoit1986asymmetry,
		skocpol1986universal,chou1987conductance,kaplan1988asymmetric,mcconville1993weak,chan1995ballistic,moon1997observation,scheffold1998universal,hoadley1999experimental,
		ghosh2000universal,trionfi2004electronic,trionfi2005time,berger2006electronic,morozov2006strong,gorbachev2007weak,heersche2007bipolar,lien2011temporal,li2012two,
		pal2012direct,matsuo2013experimental,shamim2014spontaneous,islam2018universal,rashidi2023universal} across a wide range of systems. Despite this comprehensive research, to the best of our knowledge, UCF has yet to yield any practical applications. Here, we propose leveraging this widely observed mesoscopic transport phenomenon to implement quantum RC. 
	
	We consider the UCF effect in a disordered mesoscopic conductor, which can be tuned by multiple gate voltages, as illustrated in Fig.~\ref{Experimental device and G}(a).
	The UCF can be modeled as a combination of free propagation regions, randomly distributed impurities, and a set of gate voltages, each described by a corresponding scattering matrix~\cite{cahay1988conductance}. 
	Although the physical effects of the three factors cannot be strictly separated and often involve complex interplay, it can effectively captures the key features of the UCF and provides a sufficient basis for illustrating quantum RC. The scattering matrix $S$ that connects the incoming ($a^+, b^-$) and outgoing ($b^+, a^-$) states illustrated in Fig.~\ref{Experimental device and G}(a) is expressed as
	\begin{align}
		S=
		\left(
		\begin{matrix}
			t & r' \\
			r & t'
		\end{matrix}
		\right),
	\end{align}
	where $t, t'$ and $r, r'$ are the transmission and reflection matrices, respectively,
	with their dimension $M$ being the number of propagating modes. The scattering probability between two given modes is the squared magnitude of the corresponding element in $S$.
	
	The free propagation of electrons is captured by the diagonal matrix as
	$S_n^p=\text{e}^{\text{i}k_id_n}\mathbb{I}$,
	with $\mathbb{I}$ the identity matrix. It describes the phase accumulation during free propagation in the $i$-th mode, where $k_i$ denotes the magnitude of the wave vector and $d_n$ is the length of the $n$-th free propagation region. Assuming identical scattering properties, each impurity is characterized by a scattering matrix
	$S_n^s=\mathbb{I}+\beta \mathbb{E}$,
	where $\mathbb{E}$ is the matrix of ones, and $\beta=(\text{e}^{\text{i}\theta}-1)/2M$. The matrix $S_n^s$ is related to the elastic mean free path $l_{el}$ through $\theta=2\sin^{-1}{\sqrt{\frac{M}{1+l_{el}}}}$, which describes that each incoming mode is scattered into all other outgoing modes
	with an equal probability $|\beta|^2$~\cite{cahay1988conductance}. In the UCF regime, the conductance $G$ exhibits extreme sensitivity to various physical parameters due to intricate interference effects, which serve as the primary resource for the quantum RC introduced here. We propose leveraging gate-voltage-induced UCF to achieve a fully electrical implementation of quantum RC.
	The main effect due to the gate voltage is introducing a phase shift $\delta\phi_V=eVt/\hbar=\delta kl$ to the wave function, where $V$ is the gate voltage, $\delta k$ is the momentum shift by gating, and $l$ represents the length of the gating region. The corresponding scattering matrix can be expressed as
	$S_n^v=\text{e}^{-\text{i}\delta k_il}\mathbb{I}$. For two scatterers in series described by $S_1$ and $S_2$, the whole scattering matrix $S_{12}=S_1\otimes S_2$ is constructed in a standard way as
	$t_{12}=t_2(I-r_1'r_2)^{-1}t_1$, 
	$r_{12}=r_1+t_1'r_2(I-r_1'r_2)^{-1}t_1$, 
	$t_{12}'=t_1'(I-r_2r_1')^{-1}t_2'$, 
	$r_{12}'=r_2'+t_2(I-r_1'r_2)^{-1}r_1't_2'$,
	where the subscripts of the elements follow those of the matrices.
	The overall $S$ matrix is then obtained by combining all scattering matrices $\{S_n^p, S_n^s, S_n^v\}$ successively. Extracting 
	the transmission matrix $t$ from $S$, the conductance of the sample is calculated by the Landauer formula as
	\begin{align}
		G=\frac{2\text{e}^2}{h}\text{Tr}(t^\dagger t).
	\end{align}
	
	The UCF serve as a promising candidate for a quantum reservoir. In a typical RC, the temporal input $u(t)$ is usually divided into encodings of identical size, like `00', `01', `10' and `11' for two-element encodings. In the present case, the encodings determine the number and values of the gate voltages applied to the sample. An efficient reservoir is expected to distinguish each encoding by generating a distinct output, which ultimately forms the internal state $w(t)$ after the complete use of $u(t)$. Consider a scenario where two gate voltages are applied to a sample, as shown in Fig.~{\ref{Experimental device and G}}(a), with the voltages set to either 0 or a fixed value $V$ to represent `0' and `1', respectively. Each encoding is then mapped into a conductance value, and we can select the current $I$ in the linear response regime, an easily measurable observable, as the output in experiments. If each encoding corresponds to a unique and well separated conductance value, the sample satisfies the requirement for the RC. As illustrated in Fig.~{\ref{Experimental device and G}}(b), one can see that this requirement is fulfilled across a wide range of $V$, demonstrating the excellent performance and wide tunability of our quantum reservoir.
	
	\section{Results}
	\label{section:results}
	\subsection{Spoken-digit recognition}
\begin{figure}[tb]
		\begin{center}
			\includegraphics[width=8.6cm]{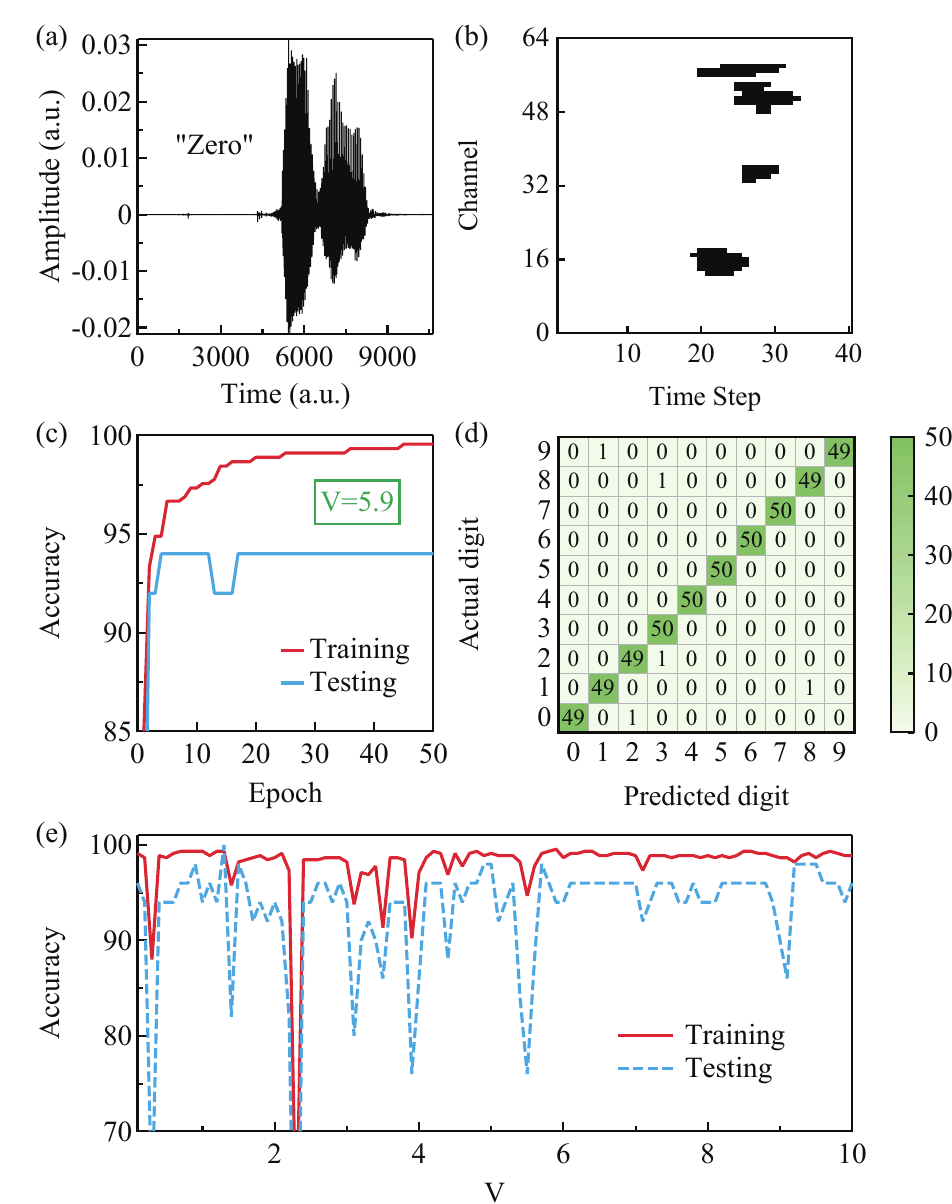}
		\end{center}
		\caption{(a) Audio waveform of digit 0 from the NIST TI46 database. (b) Cochleagram with 64 frequency channels and 40 time steps, converted from (a) using Lyon’s passive ear model and setting a threshold. (c) Training and testing accuracy as functions of iterations when $V=5.9$. The highest training and testing results are 99.6$\%$ and 94$\%$, respectively. (d) Confusion matrix visualizes the distribution of predicted digits compared to the actual spoken digits when $V=5.9$. (e) Training and testing accuracy as functions of $V$.}
		\label{spoken-digit recognition}
	\end{figure}

	To validate our quantum RC scheme, we perform two standard benchmark tasks: spoken-digit recognition and time-series forecasting. In the first task, we aim to distinguish spoken digits by ``listening" to sounds in the NIST TI46 database \cite{database}, which provides audio waveforms of isolated digits (0–9 in English) pronounced by five different female speakers. Following standard speech recognition procedure, the original audio waveform (an example for digit ``zero'' shown in Fig.~{\ref{spoken-digit recognition}}(a)) is transformed into a graph with 64 frequency channels and 40 time steps using the Lyon’s passive ear model \cite{lyon1982computational}. The graph is then converted into a digital cochleagram by setting a threshold, as illustrated in Fig.~{\ref{spoken-digit recognition}}(b), which serves as the input matrix by mapping the black and white pixels into ‘1’ and ‘0’, respectively. 
	{Directly processing the entire input requires applying $64\times 40$ gates to the mesoscopic conductor, resulting in $2^{64\times 40}$ UCF curves, which are practically indistinguishable.} 
	Therefore, we divide the inputs into smaller sections. Here, each section consists of 4 pixels in a row, which represents an input encoded by 4 gates and produces an output through quantum transport. A quantum RC composed of 640 such samples is used to process these $64\times 10$ encodings, creating a $640\times 1$ internal state to realize spoken-digit recognition by a single-layer MATLAB neural network code. 
	
	We obtain 500 cochleagrams from the NIST TI46 database, which are split into a training set of 450 and a testing set of 50. Setting all 4 gate voltages to $V=5.9$, the training result achieves a high recognition rate of $99.6\%$ within about 50 epochs. A similar performance is observed for the testing set with $94\%$ accuracy in 10 epochs, as exhibited in Fig.~{\ref{spoken-digit recognition}}(c). The confusion matrix in Fig.~{\ref{spoken-digit recognition}}(d) visualizes the distribution of predicted digits compared to the actual spoken digits. These figures demonstrate the effectiveness and fast training speed of our reservoir. Furthermore, the impact of different gate voltages $V$ on training and testing accuracy is explored in Fig.~{\ref{spoken-digit recognition}}(e), which highlights the wide tunability of the quantum RC. Notably, over 80$\%$ of the training accuracy exceeds 98$\%$, and approximately half of the testing accuracy exceeds 95$\%$. A second sample with a different disorder configuration is also used for calculation under four gates, showing similar results and further confirming the reliability of our scheme. 
	For a sample with five gates, more precise voltage adjustments are required to achieve optimal accuracy due to the limited resolution among 32 encodings. Additional details are provided in Appendix~\ref{appendix:spoken-digit recognition}.

	\subsection{Time-series forecasting}
	The second practical task is forecasting time-series of a second-order nonlinear dynamical system, the NARMA2 system \cite{atiya2000new}, described as
	\begin{align}
		y(t)=0.4y(t-1)+0.4y(t-1)y(t-2)+0.6u^3(t)+0.1.
		\label{NARMA2}
	\end{align}
	The output $y(t)$ depends not only on the current input $u(t)$ but also on the past two outputs, requiring the quantum reservoir to have short-term memory. The overall input is generated by randomly creating 40000 binary numbers, mapping `0' to 0 and `1' to 0.5, and setting $u(t)=0$ for $t<1$. The output is then calculated using Eq. \eqref{NARMA2}. Our goal is to predict the final 20000 outputs by feeding the quantum RC, trained on the first half of inputs, with the last 20000 inputs. For the training process, the first half of input is divided into 20000 overlapping sequences of four elements: $[u(-2), u(-1), u(0), u(1)], [u(-1), u(0), u(1), u(2)]$ and so on, and each encoding produces an output through a sample with 4 gates like the first task. By adjusting $V$ to 40 distinct values (referred to as size), a $40\times 20000$ internal state $w(t)$ is obtained. We introduce a weight matrix $W_\text{out}$ which transforms the internal state to the desired outputs: $y_\text{target}= W_\text{out}w$. This is a linear regression problem and the weight matrix $W_\text{out}$ is given by $W_\text{out}=y_\text{target}w^\text{T}(ww^\text{T})^+$, where the symbol `$+$' represents Moore-Penrose pseudoinverse \cite{lukovsevivcius2009reservoir}.
	
\begin{figure}[tb]
		\begin{center}
			\includegraphics[width=8.6cm]{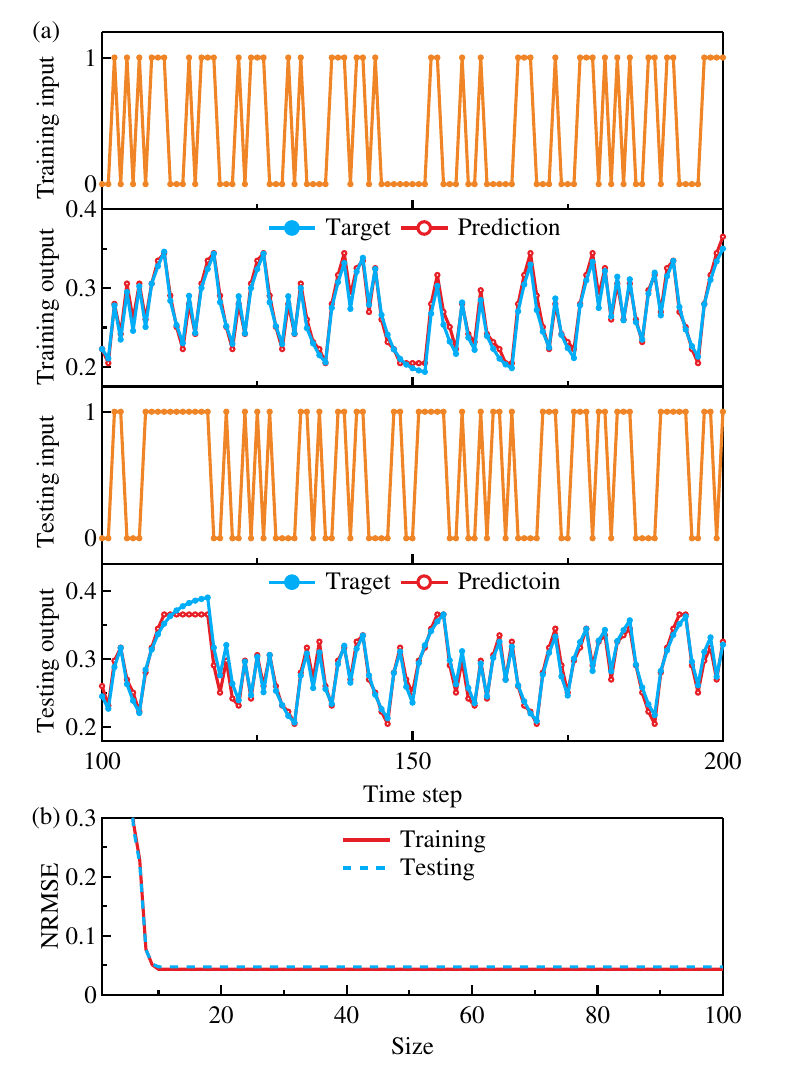}
		\end{center}
		\caption{(a) Training and testing inputs and outputs when the training process utilizes 40 distinct values of $V$. Only 100 frames are shown in these figures for clarity. (b) NRMSE trends as functions of size.}
		\label{time-series forecasting}
	\end{figure}

	As explained above, the randomly generated input is divided into training and testing sets of equal size. Fig.~{\ref{time-series forecasting}}(a) shows the training set, testing set, along with their corresponding outputs $y_{\text{target}}(t)$ and predictions $y(t)$. The normalized root-mean-square error (NRMSE) is employed to quantify the performance, which is described as
	\begin{align}
		\text{NRMSE}=\sqrt{\frac{\langle\Vert y-y_\text{target}\Vert^2\rangle}{\langle\Vert y_\text{target}-\langle y_\text{target}\rangle\Vert^2\rangle}},
	\end{align}
	where $\Vert\cdot\Vert$ denotes the Euclidean distance. A lower NRMSE indicates better performance. For this prediction, the NRMSEs are estimated to be 0.043 for training and 0.047 for testing. By adjusting the size from 1 to 100, the NRMSE saturates after its size exceeds 10, as depicted in Fig.~{\ref{time-series forecasting}}(b). 
	The normalized mean-square error (NMSE) serves as another performance metric, defined as 
	\begin{align}
		\text{NMSE}=\frac{\Vert y-y_\text{target}\Vert^2}{\Vert y_\text{target}\Vert^2}.
	\end{align}
	In our work, the training and testing NMSEs saturate at 0.0038 and 0.0042, respectively. For other quantum reservoirs, an NRMSE of 0.11~\cite{kubota2023temporal} and NMSE values of $10^{-5}-10^{-6}$~\cite{fujii2017harnessing,nakajima2019boosting} are achievable. Meanwhile, classical reservoirs attain an NRMSE of 0.0964~\cite{liu2023interface} and an NMSE of 0.020~\cite{nishioka2022edge}. Clearly, our work achieves performance comparable to other quantum reservoirs and superior to classical reservoirs, demonstrating the validity of our quantum RC. 
	It is noteworthy that performance in such tasks is influenced by differences in input settings, e.g., types of input sequences, size of training and testing sets, as well as parameter choices of network. To facilitate a fair and transparent comparison, we summarize them in Appendix~\ref{appendix:time-series forecasting}.
	Another concise way to investigate the computational power of the presented quantum reservoir is to analyze information processing
	capacity~\cite{dambre2012information,kubota2021unifying,nokkala2021gaussian,martinez2023information}.
	Further details on considering alternative sample configurations with different UCF curves using 4 and 5 gates are shown in Appendix~\ref{appendix:time-series forecasting}. The NRMSEs saturate to 0.044 in both cases.
	
	\section{Discussion and summary}
	\label{section: discussion and summary}
	In reality, the decoherence effect is inevitably present and we examine its impact within the framework of the scattering matrix approach~\cite{pala2004statistical}. Random phases are introduced in the free-propagation and gating regions, modifying the scattering matrix to $S^{\tilde{p},\tilde{v}}_n=S_n^{p,v}\otimes S_n^{\text{random}}$, where $S_n^{\text{random}}=\text{e}^{\text{i}\Delta\phi_{i,n}}\mathbb{I}$ and $\Delta\phi_{i, n}$ being the random phase fluctuation for the $i$-th mode with corresponding length $d_n$ or $l$. The random term $\Delta\phi_{i,n}$ obeys a Gaussian distribution with a mean of zero and a standard deviation of $\sigma_n=\sqrt{\{d_n,l\}/l_\phi}$. Taking into account the probabilistic nature of $\Delta\phi_{i, n}$, the conductance can be obtained through Monte Carlo averaging over a sufficiently large ensemble. As decoherence increases, the UCF curves gradually lose their characteristic structure (see Appendix~\ref{appendix:decoherence}). All UCF curves gradually overlap and become indistinguishable in the strong decoherence limit, which confirms the quantum nature of our reservoir. Therefore, our scheme becomes ineffective when the encoding differences are smaller than the resolution limit of measurement.
	
	Some remarks are made below about the experimental implementation of the gate-voltage-induced UCF, where the sample lengths $L$ typically exceed $l_\phi$. In Si metal-oxide-semiconductor field-effect-transistors, the amplitude of fluctuations can reach around $2e^2/h$ \cite{kaplan1988asymmetric} and even about $10e^2/h$ for large sample widths exceeding $l_\phi$ \cite{chou1987conductance}. 
	The UCF are also observed in both monolayer graphene, with $\sqrt{\langle\delta G^2\rangle}$ around $e^2/h$ \cite{pal2012direct}, and bilayer graphene, with amplitudes of $4e^2/h$ \cite{gorbachev2007weak}. When $L<L_\phi$, the conductance exhibits fluctuations on the order of $e^2/h$ \cite{rashidi2023universal}. Notably,
	fluctuations about $10^{-2} e^2/h$ has been reported in the topological insulator~\cite{islam2018universal} in the $L\gg l_\phi$ regime.
	This indicates that even in the presence of decoherence effects, the high measurement accuracy ensures the successful distinction of different conductance values and the implementation of quantum RC, further underscoring the universality of our approach.
	
	To implement quantum RC, the echo state property usually needs to be fulfilled~\cite{cucchi2022hands,PhysRevE.110.024207,3775-4hfd,kobayashi2024coherence}, meaning that the reservoir’s response is uniquely determined by the fading history of the input $u(t)$, independent of the reservoir’s initial conditions. This requirement rules out systems with overly strong internal dynamics, where the evolution is dictated primarily by the initial state rather than by the input. Otherwise, one would need to carefully prepare specific initial states before running quantum RC, a task that is inconvenient and particularly challenging in a quantum many-body system. In contrast to conventional quantum reservoirs, our reservoir does not suffer from initial-state uncertainty, as the propagating electrons naturally inherit the Fermi–Dirac distribution of the macroscopic electrodes. Therefore, once quantum RC is demonstrated to work for such initial states, no further concern about initial-state preparation is needed, highlighting a key advantage of our reservoir. Moreover, as shown in Fig.~{\ref{Experimental device and G}}(b), our reservoir operates in a parameter regime where distinct input encodings yield unique and well-separated conductance values. In this regime, the input signal dominates over the reservoir's internal dynamics. Consequently, our reservoir naturally fulfills the requirements for implementing quantum RC.
    
	In summary, we propose to achieve quantum RC using mesoscopic transport phenomena, and showcase how UCF can be leveraged for this purpose. The mesoscopic-based quantum reservoir offers several advantages over previous quantum RC schemes, including compatibility with existing device fabrication techniques, easily measurable output and diverse measurement approaches, and robustness to measurement back-action. Taking UCF as a specific example, low error rates and NRMSEs can be achieved for the spoken-digit recognition and time-series forecasting, respectively. Our proposal enables fully electrical implementation of quantum RC, greatly simplifying its control and manipulation. Moreover, it operates at the nanoscale, much smaller than other physical systems, which is advantageous for on-chip quantum RC and may facilitate applications with large-scale integration. Our work not only opens up a new avenue for practical quantum RC, but may also provide innovative solutions to various challenges in mesoscopic quantum information science.
	
	\begin{acknowledgments}
		W.C. acknowledges financial support from the Natural Science Foundation of Jiangsu Province (No. BK20250008 and BK20233001),
		the National Natural Science Foundation of
		China (No.  12222406), 
		the Fundamental Research Funds for the Central Universities (No. 2024300415), and
		the National Key Projects for Research and Development of China (No. 2022YFA120470).
		S.L. acknowledges financial support from the National Key R$\&$D Program of China (No. 2023YFF1203600), the AI $\&$ AI for Science Project of Nanjing University (No. 14380240), the Fundamental Research Funds for the Central Universities (No. 14380227, 14380247, and 14380250).
	\end{acknowledgments}
	
	\appendix
	\section{More details on spoken-digit recognition}
	\label{appendix:spoken-digit recognition}
\begin{figure}[htpb]
		\begin{center}
			\includegraphics[width=8.6cm]{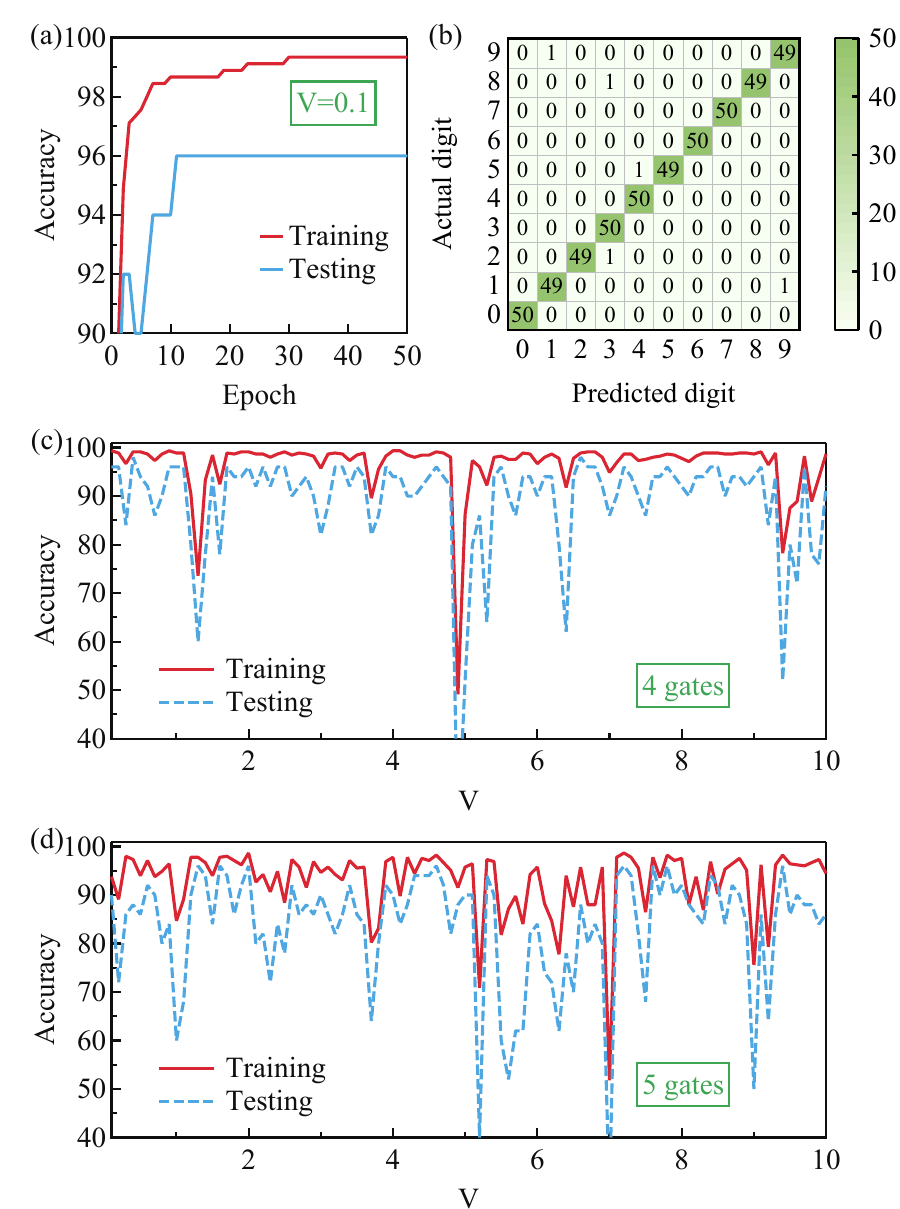}
		\end{center}
		\caption{
			For the second sample with four gates: (a) Training and testing accuracy as functions of iterations when $V=0.1$. The highest training and testing results are 99.3$\%$ and 96$\%$, respectively. (b) Confusion matrix visualizes the distribution of predicted digits compared to the actual spoken digits when $V=0.1$. (c) Training and testing accuracy as functions of $V$. For the third sample with five gates: (d) Training and testing accuracy as functions of $V$.}
		\label{spoken}
	\end{figure}

	A second sample having a different disorder configuration (different UCF curves) from sample in the main text is also used for calculation under four gates. The highest success rates reach 99.3$\%$ and 96$\%$ for training and testing in 30 and 10 epochs when $V=0.1$, respectively, as depicted in Fig.~{\ref{spoken}}(a). Fig.~{\ref{spoken}}(c) demonstrates the wide adjustability of the quantum RC. Adjusting $V$, more than 60$\%$ of the training accuracy is no less than 98$\%$, and 63$\%$ of the testing testing accuracy exceeds 95$\%$. All these data further confirm the reliability of our scheme. Finally, we consider another sample with five gates, where the inputs are divided into sections consisting of 5 pixels in a row. Adjusting $V$, only 5$\%$ reach 98$\%$ for training, and 8$\%$ reach 95$\%$ for testing, which arise from the insufficient resolution among 32 encodings, as illustrated in Figs.~{\ref{spoken}}(d). It is still usable, but we need to adjust the voltage more carefully to achieve the highest accuracy.
	
	\section{More details on time-series forecasting}
	\label{appendix:time-series forecasting}
\begin{figure}[htpb]
		\begin{center}
			\includegraphics[width=8.6cm]{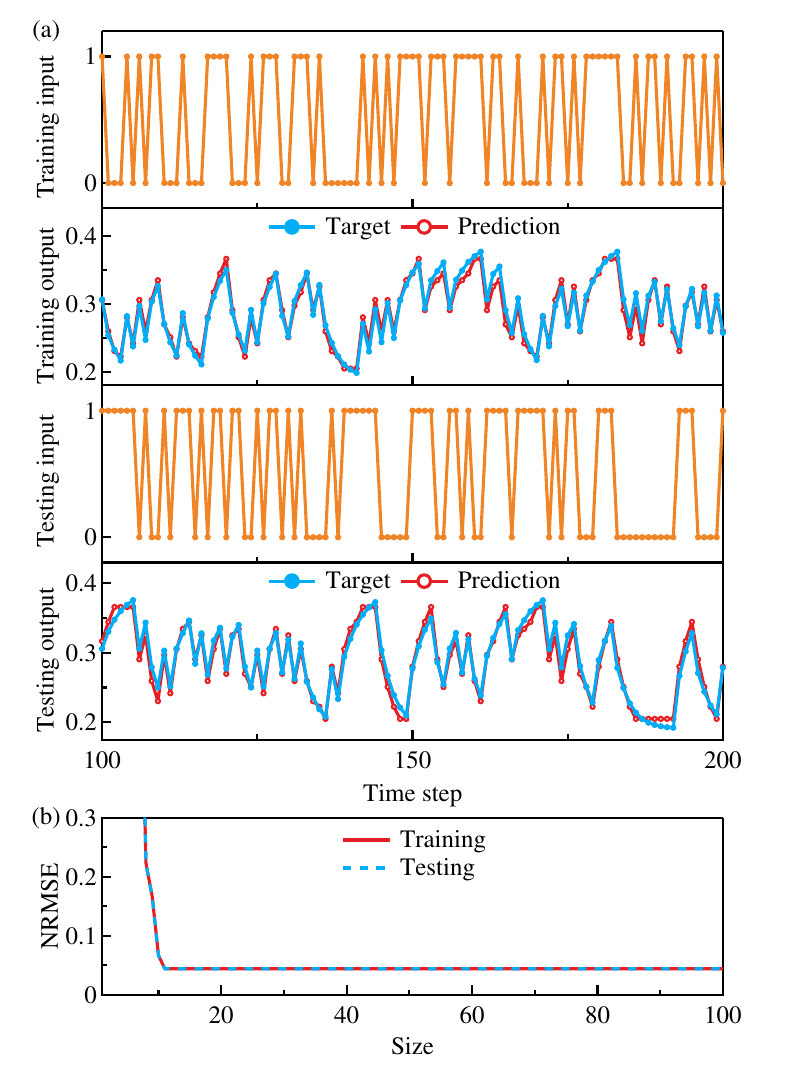}
		\end{center}
		\caption{For the second sample with four gates: (a) Training and testing inputs and outputs when the training process utilizes 40 distinct values of $V$. Only 100 frames are shown in these figures for clarity. (b) NRMSE trends as functions of size.}
		\label{4_3}
	\end{figure}

	A second sample with four gates but distinct UCF curves is used for calculation. Initially, we also train it using only 40 values of $V$, as shown in Fig.~{\ref{4_3}}(a). As depicted in Fig.~{\ref{4_3}}(b), the NRMSEs (NMSEs) are 0.044 (0.0038) for both training and testing cases. After using more values of $V$, the NRMSEs saturate after size exceeds 10. Additionally, a sample with five gates is used, which exhibits slightly worse performance compared to samples with four gates. The NRMSEs (NMSEs) saturate to 0.044 (0.0038) when size exceeds 20, which is attributed to the insufficient resolution among 32 encodings as we discussed above. These results further confirm the reliability of our scheme. In Table {\ref{table}}, we summarize several reported reservoir systems and their performance in NARMA2 tasks compared with our work.
	
	\begin{figure}[htpb]
		\begin{center}
			\includegraphics[width=8.6cm]{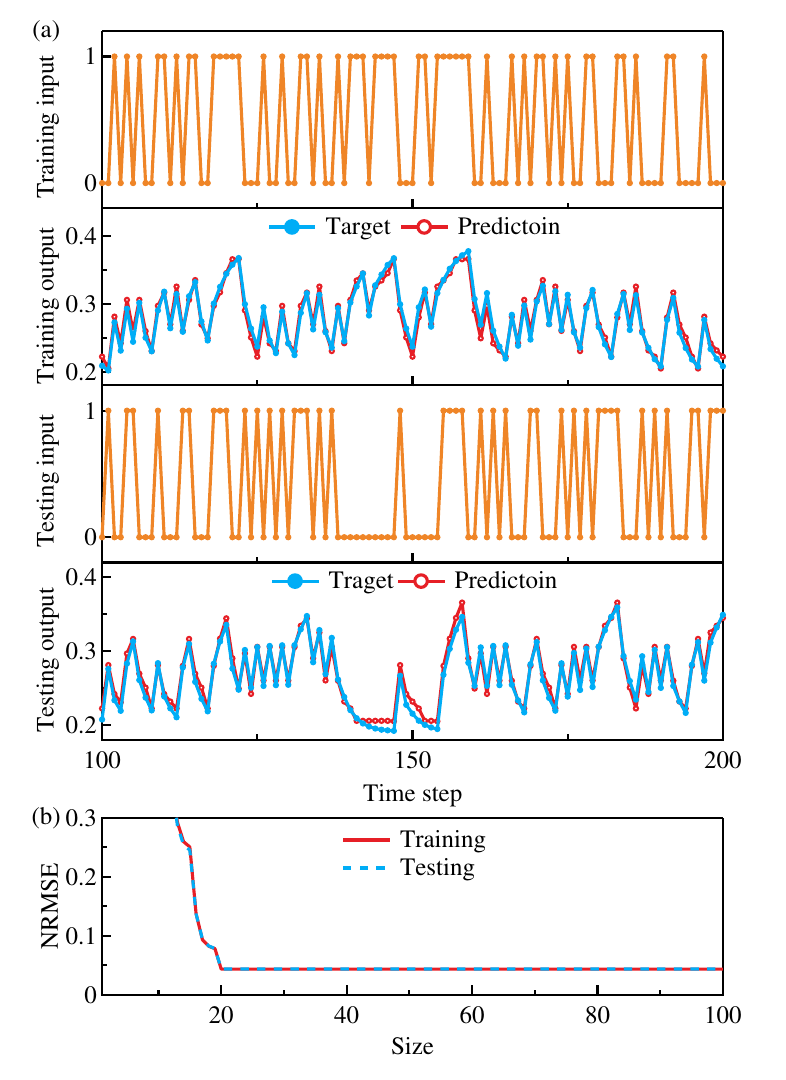}
		\end{center}
		\caption{For the third sample with five gates: (a) Training and testing inputs and outputs when the training process utilizes 40 distinct values of $V$. Only 100 frames are shown in these figures for clarity. (b) NRMSE trends as functions of size.}
		\label{5_1}
	\end{figure}
	
	\begin{table*}[pht]
		\centering
		\caption{We compare several reported reservoir systems and their performance in NARMA2 tasks with our work. The top three rows summarize the performance of quantum RC. Inputs are randomly selected from a set (using $\{,\}$) or an interval (using [,]). N/A indicates metrics that are Not Available.}
		\begin{tabular}{c c c c c c}
			\hline\hline
			Devices & Input & Training data size & Testing data size & NRMSE & NMSE \\ \hline
			Our work & $\{0,0.5\}$ & 20000 & 20000 & 0.043 & 0.0038\\
			Nuclear-magnetic-resonance spin ensemble system \cite{fujii2017harnessing,nakajima2019boosting}
			& [0,0.2] & 2000 & 2000 & N/A & $10^{-5}-10^{-6}$ \\
			IBM quantum platforms \cite{kubota2023temporal} & [0,1] & 20000 & 20000 & 0.11 & N/A \\
			Memristors \cite{du2017reservoir} & [0,0.5] & 50 & 250 & N/A & $3.61\times10^{-3}$\\
			Ion-gating reservoir \cite{nishioka2022edge} & [0,0.5] & 450 & 150 & N/A & 0.020 \\
			HZO-based interface ion dynamic transistor \cite{liu2023interface} & N/A & less than 1000 & less than 1000 & 0.0964 & N/A\\
			\hline\hline
		\end{tabular}
		\label{table}
	\end{table*}
	
	\section{UCF under decoherence}
	\label{appendix:decoherence}
\begin{figure}[H]
		\begin{center}
			\includegraphics[width=8.6cm]{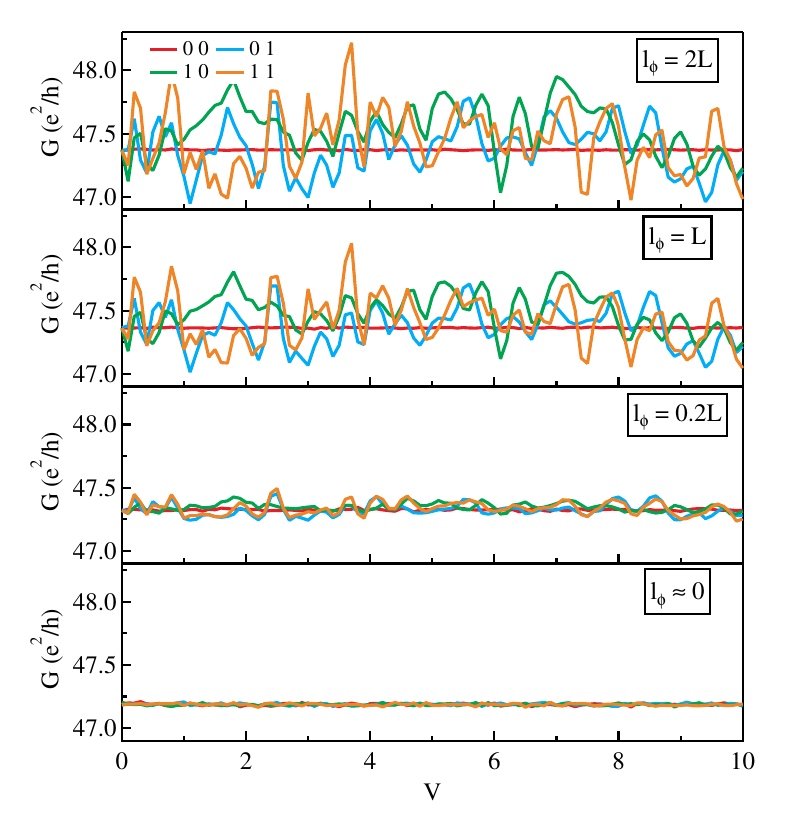}
		\end{center}
		\caption{Conductance after considering decoherence. The $l_\phi$ decreases from top to bottom and an ensemble contains 2000 samples. Other conditions are the same as Fig.~\ref{Experimental device and G}.}
		\label{UCF under decoherence}
	\end{figure}

	Following the discussion in the main text, we examine the impact of decoherence within the framework of the scattering matrix.
	We calculate the UCF for various phase coherence lengths: $l^1_{\phi} = 2L$, $l^2_{\phi} = L$, $l^3_{\phi} = 0.2L$ and $l^4_{\phi} \approx 0$, where $L=\sum_nd_n+2l$ is the sample length. 
	Other conditions are the same as those in Fig.~\ref{Experimental device and G}(b).
	The conductance under decoherence can be obtained through Monte Carlo averaging over a sufficiently large ensemble of 2000 systems with random $\Delta\phi_{i, n}$.
	For identical simulation conditions, the average and maximum deviations between different runs are less than 0.01 $e^2/h$ and 0.03 $e^2/h$, indicating adequate statistical sampling.
	As shown in Fig.~{\ref{UCF under decoherence}}, as decoherence increases, the UCF curves gradually lose their structure and become flatter. 
	All UCF curves almost overlap in the strong decoherence limit, which confirms the quantum nature of our reservoir. 

\end{document}